\documentclass[%
 aip,
 amsmath,amssymb,
reprint,%
]{revtex4-1}

\usepackage{graphicx}
\usepackage{dcolumn}
\usepackage{bm}

\usepackage[utf8]{inputenc}
\usepackage[T1]{fontenc}
\usepackage{mathptmx}
\usepackage[dvipsnames]{xcolor}
\begin{document}

\preprint{AIP/123-QED}

\title[]{Electrically tuneable exciton energy exchange between spatially separated 2-dimensional semiconductors in a microcavity}

\author{H.A. Fernandez}
\author{F. Withers}
\author{S. Russo}
\author{W.L. Barnes}
\email{w.l.barnes@exeter.ac.uk}
\affiliation{Department of Physics and Astronomy, University of Exeter, Exeter, United Kingdom, EX4 4QL}

\date{\today}

\begin{abstract}
Electrical control over the energy exchange between exciton states mediated by cavity-polaritons at room temperature is demonstrated.
Spatially separated field effect transistors based on monolayers of WS$_2$ and MoS$_2$ are placed in a tuneable Fabry-P\'erot microcavity.
This device is specially designed for the formation of exciton-polaritons that combine the two exciton species and a tuneable cavity mode.
It is further shown that the tuning of the free carrier density in the WS$_2$ film leads to a strong modulation of the Rabi splitting that modifies the excitonic and photonic nature of exciton-polaritons.
Electrical control of polaritonic devices may lead to technological applications using switchable quantum states.
\end{abstract}

\maketitle

\section{Introduction}
Exciton-polaritons are bosonic quasiparticles formed by the hybridization of the vacuum field with exciton states through an efficient energy exchange between the two constituents\cite{Hertzog2019,Torma2015} and they inherit a combination of properties of light and matter, such as an effective mass 10$^{-11}$ times the mass of an atom, and bosonic properties that allow for the observation of Bose-Einstein condensation at room temperature,\cite{Plumhof2014} lasing,\cite{Christopoulos2007} topological insulators,\cite{Klembt2018} and superfluidity.\cite{Amo2009}
The vacuum field can also hybridize with two different exciton species which has been observed in organic materials that are either mixed\cite{Coles2014,Zhong2016,Georgiou2018} or spatially separated\cite{Lidzey2000,Zhong2017,Georgiou2018} in an optical microcavity.
This multiple hybridization allows for an efficient energy exchange between the different exciton states mediated by polaritons and is effective for a separation between excitons that is larger than the typical length for F\"orster energy transfer.\cite{Andrew2000,Du2018}
To date, such polariton mediated exciton energy exchange has not been observed in spatially separated atomically thin excitonic materials, which would present an interesting way to provide remote control over optical and physical properties of materials through the manipulation of light-matter states.
\par
Atomically thin semiconductors, such as transition metal dichalcogenides (TMDs), offer remarkable optical and electronic properties for the study of light-matter interactions at room temperature.\cite{Vasilevskiy2015,Chen2017,Flatten2016,Flatten2017,Sun2017,Wang2016}
In monolayer form they exhibit a direct band gap at the K and K' points of the first Brillouin zone and a large binding energy of up to 400 meV that allows for an exciton-mediated absorption of up to 15\%.\cite{Li2014}
Among TMDs of chemical formula MX$_2$ (M = W, Mo; X = S, Se), WS$_2$ exhibits the largest oscillator strength of neutral excitons, which we estimated to be 2.7 from our experiments.
This large oscillator strength allows for the excitation of exciton-polaritons in the strong-coupling regime in a microcavity of moderate quality factor (Q $\approx$ 100).\cite{Flatten2016,Fernandez2019} MoS$_2$ is another stable semiconductor that supports exciton-polaritons in the intermediate-coupling regime,\cite{Chen2017} due to a smaller oscillator strength that we estimated to be 1.4.
\par
In our study we combine WS$_2$- and MoS$_2$-based field effect transistors (FET) in a tuneable Fabry-P\'erot microcavity made using 40 nm thick silver mirrors evaporated on quartz glass.
A hexagonal boron nitride (hBN) flake is used as a dielectric spacer between the silver film and the TMD.
Each TMD-based FET is placed on one mirror of the tuneable system, and we control the air gap thickness that separates the two excitonic materials with a precision of 3 nm.
In this system we perform optical transmission measurements as a function of the air gap thickness ranging from 800 to 1100 nm.
The exciton states of both materials hybridize with the confined field of the microcavity allowing for the formation of exciton-polaritons that combine the properties of the three constituents; the cavity photons, and the two exciton species supported by WS$_2$ and MoS$_2$ at wavelengths of 624 and 669 nm respectively.
We then manipulate the oscillator strength of one exciton specie through field effect gating using a method that we previously reported,\cite{Fernandez2019} and we show that this gating alters the nature of the exciton-polariton states.

\section{Results and discussion}
\begin{figure}
    \centering
    \includegraphics[width=0.34\textwidth]{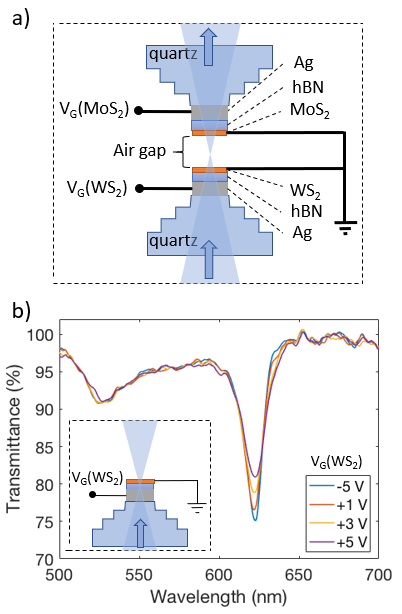}
    \caption{(a) Schematic figure of the tuneable microcavity with two transistors embedded inside. The top transistor is based on MoS$_2$, and the bottom is based on WS$_2$ as active materials. The arrows indicate the direction of propagation of the white light. The triangles indicate the white light cone of our confocal system. (b) Transmittance of the WS$_2$ monolayer for different gate voltages, from -5 to +5 V. The sample for these measurements is shown in the inset of this figure. The transmittance dip at 624 nm is associated with the WS$_2$ exciton transition.}
    \label{fig:fig1}
\end{figure}

We designed a tuneable microcavity with two TMD-based transistors embedded inside it as shown in Fig.~\ref{fig:fig1}(a).
On each mirror of the microcavity we constructed a FET based on monolayers of two different TMD materials; a WS$_2$-based FET on the bottom mirror, and a MoS$_2$-based FET on the top mirror.
The transistors consist of a stack of a monolayer TMD placed on a hexagonal boron nitride (hBN) flake, and this stack is placed onto the silver film that acts as a microcavity mirror as well as a gate electrode.
A gate voltage is applied independently to these silver electrodes while the TMD flakes are kept to ground.
The hBN flake acts as an electric insulator for the FET, and also as a dielectric spacer for the microcavity.
According to our calculations, the optimum position for the WS$_2$ to interact with the antinode of the confined electric field in the microcavity requires a thickness of the hBN flake of 60 nm, and for MoS$_2$ the required thickness is 80 nm.
In our sample, the thickness of the hBN flakes are 46 $\pm$ 2 nm for the WS$_2$-based FET, and 88 $\pm$ 2 nm for the MoS$_2$-based FET, as measured by AFM (see supplementary information, Figs.~S4 and S5).
This results in an effective Rabi splitting that is at least a 90\% of the largest Rabi splitting that would be achieved with an optimum hBN thickness.
Numerical calculations of the confined field in the microcavity using the experimental parameters of the different materials (thickness and refractive index) are shown in the supplementary information, Fig.~S3.\par
We first measured the transmittance of the WS$_2$- and the MoS$_2$ based transistors for values of the gate voltage between -5 and +5 V.
The results for the WS$_2$-based FET are shown in Fig.~\ref{fig:fig1}(b).
In this figure we observe a decrease of the transmittance dip of the A exciton transition in the WS$_2$ monolayer at a wavelength of 624 nm that is significant only for positive values of the gate voltage.
We observed no significant change in the transmittance of the MoS$_2$ monolayer for the same range of voltages (see supplementary information, Fig.~S1(b)).
As we reported previously\cite{Fernandez2019} the change in the transmittance of the WS$_2$ monolayer is attributed to an induced Coulomb screening that reduces the oscillator strength of WS$_2$ excitons due to an increase of the free charge carriers populating the conduction band of the TMD when positive voltages are applied.
Additionally the leakage current through the hBN barrier is negligible (see supplementary information, Fig.~S1(c)), which confirms that the charge carriers are being accumulated in the WS$_2$ flake.\par
\begin{figure}
    \centering
    \includegraphics[width=0.48\textwidth]{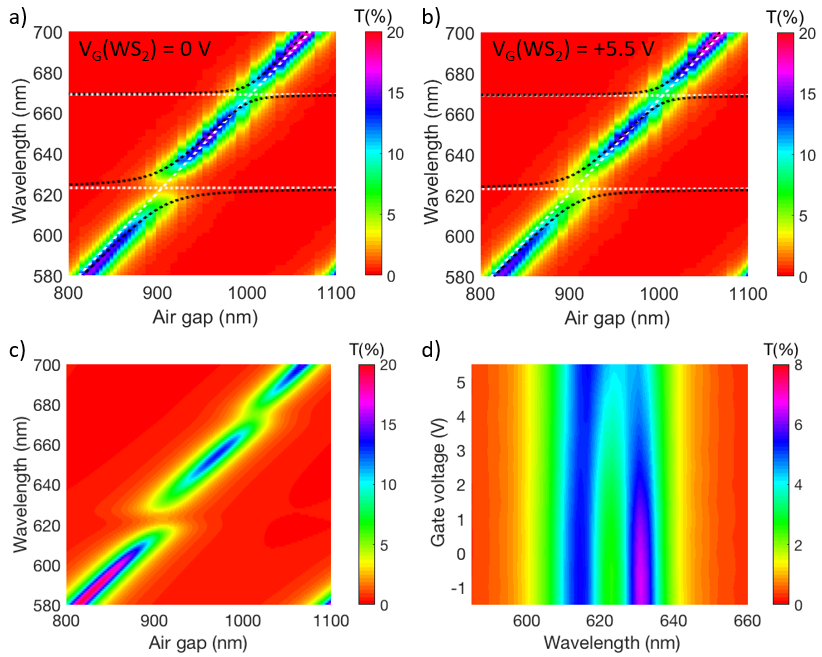}
    \caption{Measurements and modelling of the tuneable microcavity with the two transistors embedded inside: (a) is the transmittance as a function of the air gap between the two transistors, for a gate voltage of 0 V in the WS$_2$- and in the MoS$_2$-based transistors. (b) Same as in (a) but with a gate voltage of +5.5 V applied in the WS$_2$-based transistor. In these plots the white dashed horizontal lines indicate the wavelength of the respective A exciton transitions, at 669 nm is the A exciton transition of MoS$_2$, and at 622 nm is the A exciton transition of WS$_2$. (c) Modelled cavity mode as a function of the air gap for the structure shown in Fig.~\ref{fig:fig1}(a). (d) Experimental transmittance peak splitting as a function of the V$_G$(WS$_2$), for a fixed air gap of 900 $\pm$ 3 nm, at which the anticrossing between the WS$_2$ exciton transition and the cavity mode occurs.}
    \label{fig:fig2}
\end{figure}

We then measured the transmittance of the microcavity structure as a function of the air gap thickness for different values of the gate voltage applied to the WS$_2$-based FET, while keeping the gate voltage of the MoS$_2$-based FET equals to zero (V$_G$(MoS$_2$) = 0 V).
In Fig.~\ref{fig:fig2} we show results of these measurements for two different values of V$_G$(WS$_2$); (a) 0 V, and (b) +5.5 V, as indicated on the figures.
We observe that the cavity mode which varies linearly in spectral position with the air gap, is split at the values of wavelength corresponding to the WS$_2$, and the MoS$_2$ exciton transitions.
These values of the wavelength are indicated by the horizontal white dashed lines; at 624 nm for the WS$_2$ exciton transition, and at 669 nm for the MoS$_2$ exciton transition.
The oblique white dashed line indicates the bare cavity mode, and the black dashed lines are the result of the three coupled oscillator model whose Hamiltonian is shown in Eq.~(\ref{eq:3_oscillators}), where the input parameter is the experimental value of the Rabi splitting for both the WS$_2$ and the MoS$_2$ exciton-polaritons, represented in Eq.~(\ref{eq:3_oscillators}) as $\hbar \Omega$, and $\hbar \Omega'$ respectively.
\begin{eqnarray}
    \mathcal{H}=\left(
    \begin{array}{ccc}
         E_{c} & \hbar \Omega /2 & \hbar \Omega' /2 \\
         \hbar \Omega /2 & E & 0 \\
         \hbar \Omega' /2 & 0 & E'
    \end{array}
    \right),
    \label{eq:3_oscillators}
\end{eqnarray}
where $E_c$ is the fourth-order cavity mode, $E$ is the WS$_2$ exciton, and $E'$ the MoS$_2$ exciton.
In Fig.~\ref{fig:fig2}(a), the splitting of the cavity mode at the anticrossing point with WS$_2$ excitons is 57 meV, and at the position of MoS$_2$ excitons is 27 meV (see supplementary information, Fig.~S4).
In this case WS$_2$, exciton-polaritons are formed in the strong coupling regime with a Rabi splitting 1.5 times larger than both the exciton and the cavity mode line widths, while MoS$_2$ exciton-polaritons are formed in the intermediate coupling regime, with a Rabi splitting that is a 70\% of the splitting required for the strong coupling regime.
In Fig.~\ref{fig:fig2}(b) we observe that the Rabi splitting of WS$_2$ exciton-polaritons is smaller than in Fig.~\ref{fig:fig2}(a) due to the screening of the oscillator strength of WS$_2$ excitons when a positive voltage is applied.\cite{Fernandez2019}\par
To calculate the transmittance of the structure shown in Fig.~\ref{fig:fig1}(a) as a function of the air gap thickness, we first determined experimentally the permittivity of the TMDs.
The procedure involved measuring the transmittance of a bare monolayer TMD that was compared with the calculated transmittance (see supplementary information, Fig.~S2).
The calculation of the transmittance was produced after fitting the optical parameters for the Lorentz model of the permittivity of the TMDs as shown in Eg.~(\ref{eq:permittivity_excitons}).
\begin{equation}
    \varepsilon_r(\omega)=1+\sum_{j=1}^3\frac{f_j\omega_{0,j}^2}{\omega_{0,j}^2-\omega^2-i\gamma_j\omega}.
    \label{eq:permittivity_excitons}
\end{equation}
The optical parameters are shown in Tables~\ref{tab:param_WS2} and \ref{tab:param_MoS2}.
\begin{table}[h!]
\caption{\label{tab:param_WS2}Parameters for the permittivity of WS$_2$.}
\begin{ruledtabular}
    \begin{tabular}{c c c c}
    $j$ & $\hbar\omega_{0,j}$ (eV) & $f_j$ & $\gamma_j$ (eV)\\
    \hline
    1 & 1.995 & 2.70 & 0.035\\
    2 & 1.960 & 0.04 & 0.040 \\
    3  & 2.350 & 5.00 & 0.200 
    \end{tabular}
\end{ruledtabular}
\end{table}
\begin{table}[h!]
\caption{\label{tab:param_MoS2}Parameters for the permittivity of MoS$_2$.}
\begin{ruledtabular}
    \begin{tabular}{c c c c}
    $j$ & $\hbar\omega_{0,j}$ (eV) & $f_j$ & $\gamma_j$ (eV)\\
    \hline
    1 & 1.84 & 1.4 & 0.040\\
    2 & 1.80 & 0.3 & 0.080 \\
    3  & 1.96 & 2.0 & 0.090 
    \end{tabular}
\end{ruledtabular}
\end{table}
For each TMD we fit three resonances; $j=1$ is associated with the A exciton transition, $j=2$ with the negatively charged trion (A$^-$), and $j=3$ with the B exciton transition.
We then use the transfer matrix approach to calculate the transmittance of the multilayer structure as a function of the air gap thickness.
We show the result of these calculations in Fig.~\ref{fig:fig2}(c).
We observe that this calculation agrees very well with our measurements for V$_G$(WS$_2$) = 0 V.
To observe the dependence of WS$_2$ exciton-polaritons on the gate voltage we fix the cavity length to 910 $\pm$ 3 nm, and we measure the transmittance for different values of V$_G$(WS$_2$), ranging from -1.5 V up to +5.5 V.
These results are shown in Fig.~\ref{fig:fig2}(d).
The transmittance peak splitting associated with WS$_2$ exciton-polaritons decreases as we apply positive voltages.
This is a result of a decrease of the oscillator strength of WS$_2$ excitons due to the Coulomb screening of free charge carriers populating the conduction band of the TMD as we apply positive voltages.\par
\begin{figure}
    \centering
    \includegraphics[width=0.5\textwidth]{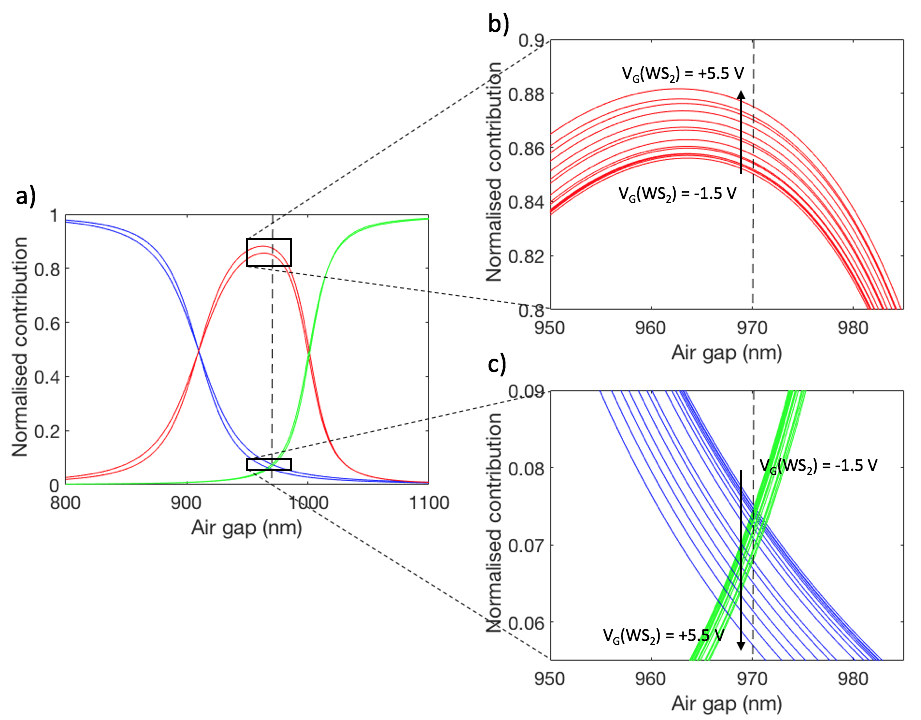}
    \caption{(a) Hopfield coefficients of the middle polariton band for two values of the Rabi splitting as a result of two values of V$_G$(WS$_2$), which are -1.5 V and +5.5 V. The blue and green lines indicate the WS$_2$ and the MoS$_2$ exciton contribution to the middle polariton band respectively, and the red lines indicate the photon contribution. The vertical dashed line indicates the value of the air gap at which the WS$_2$ and the MoS$_2$ excitons have the same contribution to the middle polariton band for V$_G$(WS$_2$) = 0, i.e. for the maximum Rabi splitting. (b) Zoom-in to the cavity photon contribution to the middle polariton band for a range of V$_G$(WS$_2$) between -1.5 V and +5.5 V. The cavity contribution increases as a more positive voltage is applied. (c) Zoom-in to the WS$_2$ and MoS$_2$ excitons contribution to the middle polariton band for a range of V$_G$(WS$_2$) between -1.5 V and +5.5 V. In this range of voltages the WS$_2$ exciton contribution decreases more rapidly than the MoS$_2$ exciton contribution.}
    \label{fig:fig3}
\end{figure}

To study the photon- and exciton-like nature of the hybrid system we quantify the contribution of the cavity photons, and the WS$_2$ and MoS$_2$ excitons to the middle polariton band of the microcavity structure.
We analyse the transmittance measurements using the three coupled oscillator model (Eq.~(\ref{eq:3_oscillators})) and we obtain the Hopfield coefficients.\cite{Hopfield1958}
In Fig.~\ref{fig:fig3}(a) we show the Hopfield coefficients of the middle polariton band of the microcavity structure with the two transistors embedded inside for different values of V$_G$(WS$_2$) while keeping V$_G$(MoS$_2$) = 0 V.
In these figures the blue lines correspond to the WS$_2$ exciton contribution to the middle polariton band, the green lines correspond to the MoS$_2$ exciton contribution, and the red line to the cavity photon contribution.
In Fig.~\ref{fig:fig3}(a) we show the WS$_2$, MoS$_2$, and cavity photon contributions as a function of the air gap thickness.
The vertical dashed line indicates the value of the air gap at which WS$_2$ and MoS$_2$ excitons contribute in the same amount to the middle polariton ban, which is 7.2\% for V$_G$(WS$_2$) = V$_G$(MoS$_2$) = 0 V.
The double lines for the three contributions are results of two different values of the $\Omega$, for gate voltages of the WS$_2$-based FET of -1.5 V and +5.5 V.\par
To elucidate this dependence of the Hopfield coefficients on $\Omega$, we analyse our measurements for a range of different values of V$_G$(WS$_2$), from -1.5 V up to +5.5 V in steps of 0.5 V.
These results are shown in Fig.~\ref{fig:fig3}(b) for the cavity photon contribution, and in Fig.~\ref{fig:fig3}(c) for the two exciton contributions.
In Fig.~\ref{fig:fig3}(b) we observe that the cavity photon contribution increases as we apply a more positive voltage, while the WS$_2$ exciton contribution decreases, as shown in Fig.~\ref{fig:fig3}(c).
We also observe a minimal decrease in the MoS$_2$ exciton contribution compared to the change in the WS$_2$ exciton contribution.\par
\begin{figure}
    \centering
    \includegraphics[width=0.35\textwidth]{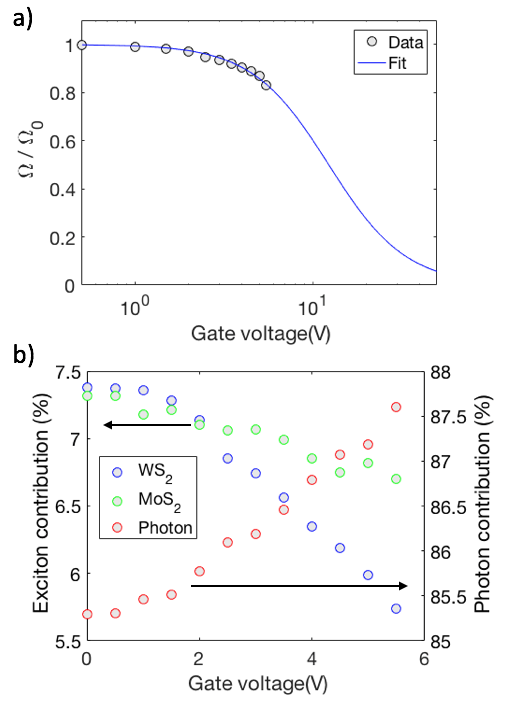}
    \caption{(a) Normalized Rabi splitting of the WS$_2$-cavity exciton-polariton as a function of V$_G$(WS$_2$). Circles are experimental data obtained from figure \ref{fig:fig2}(c). The blue line is a non-linear fit of the saturation function to the experimental data. (b) Exciton and photon contribution to the middle polariton band as a function of V$_G$(WS$_2$). These data are obtained from figure \ref{fig:fig3} (b) and (c), at a value of the air gap of 970 nm. As the gate voltage increases the WS$_2$ exciton contribution decreases rapidly and the cavity photon contribution increases in the same way.}
    \label{fig:fig4}
\end{figure}

To comprise our observations, in Fig.~\ref{fig:fig4}(a) we show the Rabi splitting of WS$_2$ exciton-polaritons normalized to the maximum Rabi splitting, which is achieved for V$_G$(WS$_2$) $\leq$ 0 V.
We fit a saturation function (Eq.~(\ref{eq:saturation})) to these data, which is shown as a blue line in Fig.~\ref{fig:fig4}(a);
\begin{equation}
    \frac{\Omega(V_G)}{\Omega_0}= \frac{1}{1+\Big(\frac{V_G}{V_G^s}\Big)^2},
    \label{eq:saturation}
\end{equation}
where $V_G^s$ is the saturation value of the gate voltage, at which $\Omega = \Omega_0$/2.
From the curve fitting procedure we obtained $V_G^s=12.3$ V.
In Fig.~\ref{fig:fig4}(b) we show the WS$_2$, and MoS$_2$ exciton contributions, and the cavity photon contribution to the middle polariton band as a function of the gate voltage applied to the WS$_2$-based FET.
These data are obtained from Fig.~\ref{fig:fig3}(c) at a fixed cavity length of 970 $\pm$ 3 nm, indicated by the vertical dashed line.
We observe a clear dependence of the WS$_2$ exciton contribution to the middle polariton band on the gate voltage.
A clear dependence is also observed in the cavity photon contribution.
As the WS$_2$ exciton contribution decreases, the cavity photon contribution increases.
There is also a small change in the MoS$_2$ exciton contribution to the middle polariton band.
\section{Summary and conclusions}
In summary, we electrically controlled the photonic and excitonic composition of the polariton bands in a microcavity system that combines two different excitonic materials.
We incorporated WS$_2$- and MoS$_2$-based FET in the microcavity, in a way that both excitonic materials hybridize with the confined field of the microcavity.\par
The cavity mode was split at wavelengths of 624 and 669 nm, corresponding to the neutral exciton transitions of WS$_2$ and MoS$_2$ respectively, providing evidence for the formation of three polariton states observed as the upper, middle, and lower polariton bands.
A splitting of 57 meV whas achieved for WS$_2$ exciton-polaritons, and 27 meV for MoS$_2$ exciton-polaritons, both formed with the fourth-order cavity mode.
We then quantified the photonic and excitonic composition of the polaritons bands by using a three coupled oscillators model, which allowed us to extract the Hopfield coefficients.
Through this analysis we found that initially the middle polariton band is composed by cavity photons in an 85.2\% and by WS$_2$ and MoS$_2$ excitons in a 7.4\% each, for a fixed value of the air gap thickness of 970 $\pm$ 3 nm.
We then control the exciton composition of the polariton bands by altering the coupling strength of WS$_2$ exciton-polaritons through field effect gating.
By applying a gate voltage we increased the free charge carrier density in the WS$_2$ monolayer, which leads to decrease of the exciton oscillator strength.
This allows for control over the WS$_2$ exciton composition of the polaritons bands.
By increasing the gate voltage, the WS$_2$ exciton composition decreased while the cavity photon composition increased.\par
The results presented here are, to the best of our knowledge, the first evidence of control over multiple hybridization of the confined field in a microcavity with different 2-dimensional semiconductors, and may lead to technological applications through switchable quantum states in optoelectronic devices.

\begin{acknowledgements}
The authors acknowledge financial support from the Engineering and Physical Sciences Research Council (EPSRC) of the United Kingdom, via the EPSRC Centre for Doctoral Training in Metamaterials (Grant No. EP/L015331/1).
FW acknowledges support from the Royal Academy of Engineering. SR acknowledges financial support from The Leverhulme Trust, research grants ``Quantum Drums" and ``Quantum Revolution".
WLB acknowledges the support of the European Research Council through the project Photmat (ERC-2016-ADG-742222: www.photmat.eu).
\end{acknowledgements}

Supplementary material is available online at aip.scitation.org/journal/apl.
Research data is available online at doi:...

\bibliography{mybib}

\end{document}